\documentclass[mathleft
]{an}
\usepackage{graphicx}
\usepackage{times}
\newcommand{\HI}{H{\,\small I}}

\newcommand{\kms}{km\ s$^{-1}$}
\newcommand{\FRI}{FR{-\small I}}

\newcommand{\sauron}{{\texttt {SAURON}}}
\begin{document}

\Pagespan{789}{}
\Yearpublication{2006}%
\Yearsubmission{2005}%
\Month{11}%
\Volume{999}%
\Issue{88}%

\title{Gas and stars in compact (young) radio sources}

\author{R.\ Morganti\inst{1,2} \fnmsep\thanks{Corresponding author:
  \email{morganti@astron.nl}\newline}
B. Emonts\inst{3}, J. Holt\inst{4}, C. Tadhunter\inst{5}, T. Oosterloo\inst{1,2} \and  C. Struve \inst{1,2}
}
\titlerunning{Gas in young radio sources}
\authorrunning{R. Morganti et al.}
\institute{
Netherlands Foundation for Research in Astronomy,
    Postbus 2, 7990 AA Dwingeloo, The Netherlands 
\and 
Kapteyn Astronomical Institute, University of Groningen
    Postbus 800, 9700 AV Groningen, The Netherlands 
\and 
Australia Telescope National Facility, CSIRO, PO Box 76, Epping, NSW 1710, Australia
\and
Leiden Observatory, Leiden University, P O Box 9513, NL-2300 RA Leiden, The Netherlands
\and
Department of  Physics and Astronomy,
University of Sheffield, Sheffield, S7 3RH, UK}

\received{1}
\accepted{1}
\publonline{later}

\keywords{Galaxies: active, Galaxies: evolution, ISM}

\abstract{
Gas can be used to trace the formation and evolution of galaxies as well as the impact that the nuclear activity has on the surrounding medium. 
For nearby compact radio sources, we have used observations of neutral hydrogen - that we detected in emission distributed over very large scales -  combined with the study of the stellar population and deep optical images to investigate the history of the formation of their host galaxy and the triggering of the activity.
For more distant and more powerful compact radio sources, we have used optical spectra and \HI\ - in absorption - to investigate the presence of fast outflows that support the idea that compact radio sources are young radio loud AGN observed during the early stages of their evolution and currently shedding their natal cocoons through extreme circumnuclear outflows.
We will review the most recent results obtained from these projects.
}

\maketitle

\section{Introduction}

The origin of the onset of the nuclear activity in  galaxies is still a matter of debate. As radio astronomers, we are in the  lucky position of knowing how to find a newly born or young radio source  (and even organise a meeting about them!) and therefore we can attempt to investigate whether the characteristics of their host galaxy can tell us about the possible trigger of the radio activity. The overwhelming majority of bright, low-$z$ ($z<1$) radio sources are hosted by early-type galaxies (ETGs).  ETGs are thought to form hierarchically, growing in mass and size by merging of smaller systems. However, it is now clear that ETGs are complex systems (see e.g.\ Emsellem et al.\ 2007). They often host sub-components in their structure and their  dynamics, stellar populations and nuclear activity  is characterised by a surprisingly wide variety. This diversity goes beyond current simulations and predictions from hierarchical models. Gas and associated dissipative processes can play an important role in explaining such a diversity. At the same time, the effects of the feedback on the gas produced by supernovae winds or by the nuclear activity is also extremely important (Silk \& Rees 1998; Hopkins et al.\  2005). Thus, {\sl the initial phase of AGN can be crucial in the evolution of the host galaxy}.

\begin{figure*}
\centering
\includegraphics[width=0.7\textwidth]{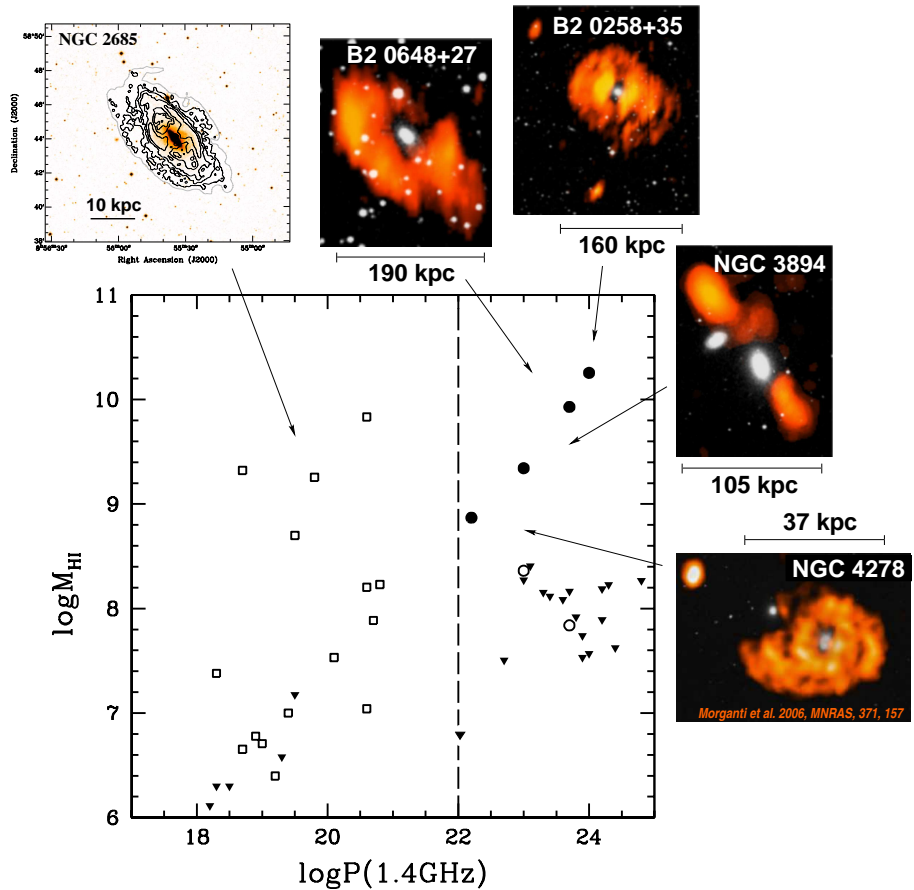}
\caption{Total \HI\ mass detected in the host galaxy plotted against the radio power for the samples of radio-loud (log $P > 10^{22}$ W Hz$^{-1}$, where P is the radio power at $v = 1.4$ GHz) and radio-quiet (log $P < 10^{22}$ W Hz$^{-1}$)early-type galaxies (from Morganti et al.\ 2006, Oosterloo et al.\ in prep.). Several galaxies with disk-like \HI\ structures are shown to clarify their morphology. The
filled circles are the detections of compact radio sources, the open circles are the detections for FRI sources while the open squares are the \HI\ detection for radio-quiet early-type galaxies.
The fill triangles are the \HI\ upper limits. 
}
\label{label1}
\end{figure*}

As extensively discussed in this meeting, GPS and CSS (Gigahertz Peaked Spectrum  and Compact Steep Spectrum) radio sources are generally considered recently started  radio sources. They are, therefore, the ideal targets to learn more about the relation between formation and evolution of the host galaxy, the trigger of the activity and its effect on the nuclear regions of the host galaxy.

In order to gain a better insight on these topics, we combine optical information of CSS/GPS sources with extensive studies of the cool neutral hydrogen (\HI) gas. The large-scale distribution and kinematics of the \HI\ gas, combined with a spectroscopic study of the stellar populations across the host galaxy, provide essential information about possible merger or interaction events that may be related to the triggering of radio sources, while the characteristics of the \HI\ and ionised gas in the central region can be used to study in great detail the total feedback that the radio sources exert on the ISM of the host galaxy. Our work  concentrates on: i) nearby, relatively low power radio sources (Fanaroff-Riley type I sources)  that we study using \HI\ detected both in emission and in absorption combined with the properties of their stellar population; ii)
more distant FRII-like radio sources for which we use,  for lack of sensitivity to detect \HI\ emission, observations of the \HI\ absorption and combine this with observations of the ionised gas.

\section{Nearby extended and compact radio sources  - presence of neutral hydrogen}

In the last few years we have searched for \HI\ around the hosts of nearby radio galaxies.  
Initial results are presented in Emonts et al.\ (2007) and Emonts (2006). 
In order to understand whether the gas content of nearby radio galaxies is at all different from that of their radio-quiet cousins, we have compared the presence of neutral hydrogen gas in these two types of objects. The aim was to determine whether e.g.\ nearby radio-loud early-type galaxies more frequently show evidence for gas-rich merger or interaction events (in the form of gaseous tidal-tails, bridges, shells or disks). 

Our study targeted a complete sample of nearby ($z \leq 0.04$) radio galaxies from the B2 catalog, consisting of relatively low radio power (log $P > 22$ W Hz$^{-1}$) \FRI\ and compact steep spectrum radio sources (see Emonts et al.\ 2007 and Emonts et al.\ in prep. for more details on the sample selection). As a comparison sample we used the \sauron\ sample studied in \HI\ by Morganti et al.\ (2006) and Oosterloo et al.\ (in prep).
In Figure 1 we plot the total \HI\ mass detected in emission versus the radio power at 1.4~GHz of the source. From Fig.\ 1  the separation in radio power of our B2 sample of radio galaxies and the \sauron\ sample is clear. Although the detection limits for the \HI\ masses of the two samples are different (due to the fact that the radio sources tend to be farther away), there is a first order similarity between the total \HI\ mass detected in radio-loud and radio-quiet early-type galaxies. Furthermore, in both cases, the high-mass end of the \HI\ detections consists of large-scale disk- and ring-like structures (with masses up to a few $\times 10^{10} M_{\odot}$ and sizes up to $\sim$200 kpc). We therefore do not find evidence in \HI\ that gas-rich mergers are more frequent among nearby low-power radio galaxies than among their radio-quiet counterparts. 

However, in terms of radio source properties we find an interesting dichotomy.  The radio galaxies from the B2 sample that contain large amounts of \HI\ gas ($> {\rm few} \times 10^{8} M_{\odot}$) {\sl all host a compact radio source}, while the more extended \FRI\ radio sources do not contain similar amounts of \HI\ (Emonts et al.\ 2007). In total about 50$\%$ of the compact sources in our B2 sample contain these large-scale \HI\ disks and/or rings. The \HI-rich compact radio sources are therefore not likely to evolve into extended \FRI\ sources as we do not see similar \HI\ structures in these radio galaxies. 
The lack of large-scale \HI\ associated with the extended \FRI\ radio galaxies means that they are likely fueled by processes other than major mergers, e.g. though minor events or by accreting their hot gaseous atmospheres (see e.g.\ Hardcastle, Evans \& Croston  2007).

\subsection{Origin of the large-scale \HI\ disks/rings}

\begin{figure*}
\centering
\includegraphics[width=0.9\textwidth]{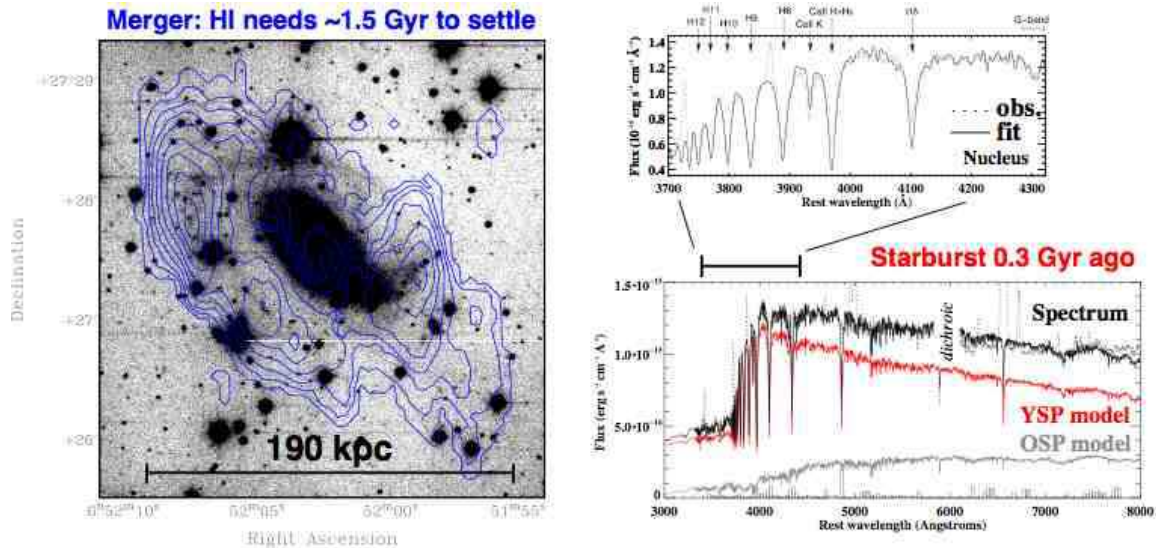}
\caption{{\sl Left:} \HI\ contours overlaid onto deep optical image of radio galaxy B2~0648+27. Contours: $0.22, 0.36, 0.52, 0.71, 0.95, 1.2, 1.5, 1.8, 2.1 \times 10^{20}$ cm$^{-2}$. {\sl Right:} results of stellar population synthesis modeling of optical long-slit spectra; the stellar light is dominated by a 0.3 Gyr young stellar population at various locations across the host galaxy. See Emonts et al. 2006 $\&$ 2008  for details.}
\label{label1}
\end{figure*}

Do the large-scale \HI\ structures that we detect around several compact radio sources tell us something about the formation history of the host galaxy? 

For one case, B2 0648+27, we studied in great detail the properties of the \HI\ gas, star formation and AGN activity (see Fig.\ 2). As discussed in detail in Emonts et al.\ (2006, 2008), the enormous \HI\ ring that surrounds the host galaxy ($M_{\rm HI} = 8.5 \times 10^{9} M_{\odot}$ and diameter = 190 kpc) was most likely formed by a merger of two gas-rich disk galaxies. This merger occurred roughly 1.5 Gyr ago, after which the \HI\ gas that was expelled in large-scale tidal features during the initial stages of the merger, had the time to fall back onto the host galaxy and settle into the regular rotating ring that we observe. Deep optical imaging reveals that a faint stellar ring follows the \HI\ gas, tracing the stellar debris that was expelled during the merger in a similar fashion as the \HI. The merger origin is confirmed by the detection of a 0.3-0.4 Gyr young post-starburst stellar population that is present throughout the host galaxy and which could have given B2 0648+27 the appearance of an (Ultra-) Luminous Infra-Red Galaxy in the first epoch after the merger. The activity in B2 0648+27 occurred in several stages, with significant time-delays between the initial merger and the starburst event, as well as between the merger/starburst and the onset of the current episode of radio-AGN activity.

The other \HI-rich compact radio sources from our B2 sample also show disks and rings of neutral hydrogen gas very similar to the case of B2 0648+27 (see Emonts 2006 for details) but that appear more settled. However, they do not show evidence for such a prominent young stellar population. Their spectra indicate that the stellar populations in the main body of these systems are at least one to several Gyr old (Emonts 2006). If these large-scale \HI\ disks/rings were also formed by major mergers, they must have happened significantly longer ago than the case of B2 0648+27. This  would also allow them to obtain their regular morphology and kinematics. 

In summary,  it is interesting to see that galaxies that appear to originate from gas-rich major mergers do not seem to evolve into an extended radio galaxy. This has implication for the evolution of the host galaxy. Possibilities are that either these radio sources  are frustrated by the ISM in the central region of the host galaxy, or  the fuelling stops before the sources can expand (see also Giroletti, Giovannini \& Taylor 2005) perhaps because of the type of merger (prograde vs retrograde, see di Matteo et al.\ 2007) that is at the  origin of these galaxies.


\section{Powerful CSS/GPS: \HI\ and ionised gas}

The above mentioned B2 sample contains relatively low power compact radio sources. More powerful GPS and CSS sources are found at higher redshift, where the sensitivity becomes too low for \HI\ emission-line studies. However, the strong radio continuum flux of GPS and CSS sources is excellent for \HI\ absorption studies.

It is known that GPS/CSS sources often contain a kinematically complex ISM, with outflows of gas seen in various phases of the warm/hot ISM (e.g.\ Gelderman $\&$ Whittle 1994, Holt, Tadhunter \& Morganti 2008 and refs. therein). As mentioned by Holt in these proceedings, we have performed a detailed  study of these outflows of ionised gas in a sample of 14 CSS/GPS radio galaxies from the 3C, 4C and 2Jy samples of radio sources (see Holt et al.\ 2008). In parallel, we study the kinematics of the \HI\ gas in the line-of-sight toward the central continuum of the GPS/CSS sources. 
From the long-slit spectra we derive detailed values of the systemic velocity (see Holt et al.\ 2008 for a full discussion). This is crucial if we want to understand how common fast gaseous outflows are and how relevant they are for the evolution of the galaxy.

An interesting result of this study is that the neutral outflows occur especially in young or restarted radio sources (Morganti, Tadhunter \& Oosterloo 2005a).  In at least
some cases we know that they are originating at few hundred pc to kpc from the nucleus, and they are most likely driven
by the interactions between the expanding radio jets and the gaseous medium
enshrouding the central regions.  We estimated that the associated mass outflow
rates are up to $\sim 50$ $M_\odot$ yr$^{-1}$, comparable (although at the
lower end of the distribution) to the outflow rates found for starburst-driven
superwinds in Ultra Luminous IR Galaxies (ULIRG), see Rupke, Veilleux \& Sanders\ (2002).
This suggests that massive, jet-driven outflows of neutral gas in radio-loud
AGN can have as large an impact on the evolution of the host galaxies as the
outflows associated with starbursts. 
The similarities found between the kinematics of the neutral and ionised gas
indicate that the two phases of the gas are part of the same outflow.
However, the outflows of ionised gas are typically much less massive (Morganti
et al.\ 2005b, Emonts et al.\ 2005) than those of the neutral hydrogen.

The GPS source 4C12.50 is a nice example of radio source showing a massive outflow of \HI\ gas. For 4C 12.50, a broad and blueshifted \HI\ absorption covers a total velocity range of almost 2000 \kms\ (see Morganti et al.\ 2005a), comparable to an outflow of ionised gas found by Holt, Tadhunter \& Morganti  (2003). A VLBI study of the neutral
hydrogen in the nuclear regions of this object shows that the narrower \HI\
component (detected close to the systemic velocity) is associated with an
off-nuclear cloud ($\sim 50$ to 100 pc from the radio core. Thus, 4C12.50 is an example of a young radio source with nuclear regions that
are enshrouded in a dense cocoon of gas and dust. The radio jets are expanding
through this cocoon, sweeping material from the nuclear regions.

An other interesting case is PKS 1549--79, discussed in great detail in Holt et al.\ (2006). Although this system does not display an outflow of neutral hydrogen, \HI\ absorption is detected and resolved against the extended radio structure at VLBI scales. In this case the \HI\ absorption likely traces a quiescent cocoon of \HI\ that surrounds the compact radio jets and which has not yet been perturbed by the radio source. We conclude that PKS 1549--79 is a radio source in a stage where the nucleus is
still hidden (in the optical) by the gas/dust coming from the merger that
triggered the radio source.  Young, small scale radio jets are expanding
through dense cocoon, sweeping aside gas and dust. Interestingly, both these sources show evidence of a young stellar population component in their optical spectra, consistent with the scenario described above.

\section{Summary and conclusions}

We used detailed observations of gas and star formation in order to trace the formation history and evolution of the host galaxies of nearby compact radio sources. Observations of large-scale \HI\ emission-line gas reveal that about half of lower power compact radio sources at low redshift contain large amounts of \HI\ gas. The gas is distributed in large-scale disk- and ring-like structures, which could be the result of either recent or old merger events. Extended \FRI\ sources do not contain similar large amounts of \HI, hence these \HI-rich low-power compact sources are not expected to grow into extended \FRI\ sources. For the lowest redshift systems as a whole, no clear difference between radio-loud and radio-quiet early-type galaxies is seen. More powerful GPS and CSS sources (which are too far away for \HI\ emission-line studies) often show evidence for massive outflows, both in the ionised gas and in the \HI\ absorption-line gas. This indicates that GPS and CSS sources, whether frustrated or not, are a vital component in the feedback of galaxies.


\end{document}